# Title: Peer assessment enhances student learning

**Authors:** Dennis L. Sun[a], Naftali Harris[a], Guenther Walther[a], and Michael Baiocchi[b]

**Affiliations:**
[a] Department of Statistics, Stanford University, Stanford CA 94305, USA.
[b] Prevention Research Center, Stanford School of Medicine, Stanford CA 94305, USA.

**Abstract:**

Feedback has a powerful influence on learning, but it is also expensive to provide. In large classes, it may even be impossible for instructors to provide individualized feedback. Peer assessment has received attention lately as a way of providing personalized feedback that scales to large classes. Besides these obvious benefits, some researchers have also conjectured that students learn by peer assessing, although no studies have ever conclusively demonstrated this effect. By conducting a randomized controlled trial in an introductory statistics class, we provide evidence that peer assessment causes significant gains in student achievement. The strength of our conclusions depends critically on the careful design of the experiment, which was made possible by a web-based platform that we developed. Hence, our study is also a proof of concept of the high-quality experiments that are possible with online tools.

**Introduction:**

Feedback is one of the single most important factors influencing student learning.[1,2] In practice, it is often not possible to provide feedback that is both detailed and prompt, limiting its effectiveness.[3] In large college classes and massively open online courses (MOOCs), providing personalized feedback to students is especially challenging. We call this the *problem of feedback*.

Although machines are now able provide automated feedback for many kinds of questions,[4] some questions still elude even the most powerful of machines. For example, consider the following open-ended question:

> Josh flips a coin 100 times. The coin comes up heads 60 times. He calculates the p-value to be about 2% for testing the null hypothesis that the coin is fair. Explain what this 2% means in the context of this problem.

The correct answer is that the 2% represents the probability of observing so many heads if the coin were fair. However, a common misconception among students is that it represents the probability the coin is fair. Even with state-of-the-art semantic parsing, machines cannot accurately discriminate incorrect answers from correct ones.[5] On the other hand, a human who understands the concept would have little difficulty distinguishing the two.

Crowdsourcing is a general strategy for tackling problems that are infeasible for a machine but relatively easy for a human. The basic idea is to leverage the power of numbers by aggregating small contributions from many humans, and has already been successfully applied to complex tasks from protein folding to character recognition.[6,7] The problem of feedback can be "crowdsourced" by having students grade one another, a practice known as *peer assessment*. Peer assessment distributes the responsibility of feedback among the hundreds—perhaps even thousands—of students in a class. It enables faster feedback in classes where instructor grading is feasible,[8] and it makes feedback possible in classes so large (e.g., MOOCs) that evaluation would otherwise be impossible.[9]

Instructors often have two main concerns about peer assessment. The first is whether students can be trusted to grade accurately. This question has been extensively studied in the literature, and the consensus is that peer grades are comparable to instructor grades.[9,10,11] The other concern is implementation; traditionally, peer assessment involved a back-and-forth exchange of papers between students. Fortunately, this labyrinthine system is a relic of the past, made obsolete by the Internet. Web-based peer assessment tools that automatically collect and distribute student responses are now available with most learning management systems (LMS). These online systems also anonymize students and graders, obviating any privacy concerns associated with peer assessment.[12]

We have seen that peer assessment solves the problem of feedback, reducing the burden on instructors without sacrificing quality. To top it all, it is often claimed that students also learn through peer assessing. If true, then peer assessment should be viewed not only as an administrative tool, but also as a pedagogical device.[8,13] Unlike most proposals for improving student learning, however, peer assessment requires minimal investment from instructors and may even save them time.

The pedagogical benefits of peer assessment have been trumpeted both in the research literature[13] and in the public sphere. Peer assessment even received national attention in 2002, when the U.S. Supreme Court issued its ruling in *Owasso v. Falvo*, a case which examined whether peer assessment violated student privacy. In his majority opinion, Justice Anthony Kennedy argued,

> Correcting a classmate's work can be as much a part of the assignment as taking the test itself. It is a way to teach material again in a new context, and it helps show students how to assist and respect fellow pupils.[12]

Unfortunately, the empirical evidence has not kept pace with the claims, with most studies based on surveys of student and teacher perceptions.[8,14,15] Only a few studies have attempted to quantify the effect on an objective criterion such as achievement, and of these, most have been correlational studies. A representative study in this category is Gerdeman et al.,[16] which examined whether peer assessment improved students' writing skills. This study lacked a control group, so it is impossible to know whether students improved any more under peer assessment than they would otherwise. Furthermore, the study measured achievement using the students' own peer grades, rather than an objective measure (e.g., scores given by a third-party observer who was blinded to the treatment). Sadler and Good[17] conducted a randomized experiment to study the effect of peer assessment on achievement, but their study lacked statistical power to make a definitive conclusion. This gap in the literature has been noted by several researchers, who have identified this as an important question for research.[17,18]

**Materials and Methods:**

We were interested in whether peer assessment could aid conceptual understanding and problem solving, two skills that are relevant in science, technology, engineering, and mathematics (STEM) classes. To investigate this, we conducted a randomized controlled trial in a large introductory statistics class.

Under the peer assessment treatment, students provided scores and comments on the homework responses of three peers, and in turn, received feedback on their own response from three peers. All homework responses and peer assessments were submitted through an online system, and submissions were anonymized before distribution. The control subjects also submitted homework responses online, but did not participate in peer assessment and had their homework graded by instructors.

To enhance the precision of the study, we employed a crossover design, meaning that every student participated in both the treatment and control groups at different points in the course. Since students can now be compared against themselves, this design effectively accounts for all difference between students, which is arguably the largest source of variability in educational studies. To do this, we divided the course into four main units. Students were randomized to four treatment arms, each one receiving treatment (T) and control (C) in a different order over the four units (TCTC, CTCT, TCCT, and CTTC).

To measure achievement, each unit was concluded by a quiz. These quizzes were designed to measure the short-term effect of peer assessment. The students also took a comprehensive final exam that measured longer-term learning. Instructors graded all assessments to ensure consistency; these instructors were blinded to the treatment groups of the students.

To further ensure against baseline differences between the treatment and control groups within a given unit, we augmented this design with matched-pairs randomization. Students were matched into the pairs based on covariate data (e.g., class year, previous statistics experience), and each pair was assigned to complementary treatment arms (e.g., if one student was assigned to TCTC, his or her pair would be assigned to CTCT). This ensures that within a given unit, every student in the treatment group will be balanced by a "similar" student in the control group. Figure 1 shows the result of the pairing. The covariate balance for the actual randomization appears in the supplement and confirms that the matched-pairs design produced balanced treatment arms.

Finally, we conducted a full replication of the study in a different academic term with a different instructor. In all, 148 students participated in the study during the first term (autumn) and 239 students during the second (winter).

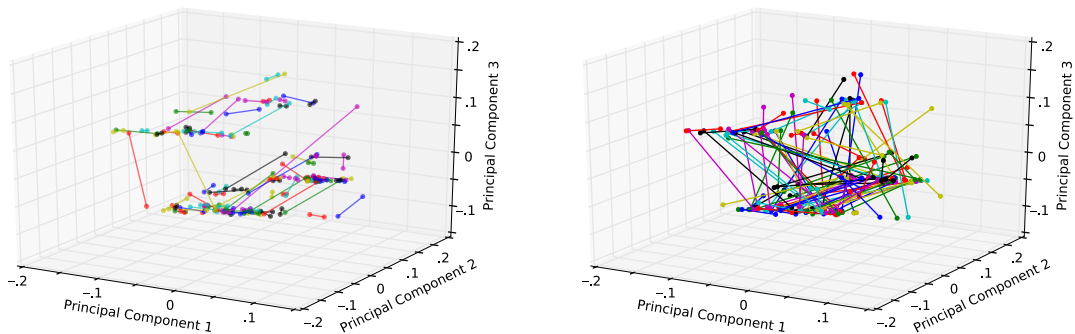

**Figure 1.** Two plots showing the effect of the matched pairs randomization design (left) as compared with complete randomization (right). Each point represents a student's covariate data, projected onto the top three principal components. A connecting edge indicates that those students have been paired and assigned to complementary treatment arms. The edges in the matched pairs design are much shorter than under complete randomization, indicating that matching produces more similar randomizations.

**Results:**

Students who participated in peer assessment during a given unit performed significantly better on the unit quizzes (Cohen's d = .115, t(298) = 2.92, p = .002) than students who did not. These students also did better on the corresponding questions on the final exam (d = .12, t(319) = 3.03, p = .001), suggesting that the benefits of peer assessment persist over time. In the context of our quizzes and exams, where the standard deviations ranged from 15 to 25 percentage points, this translates to a 2 to 3 percentage point increase in the average score. Figure 2, which depicts the actual distribution of scores for one of the unit quizzes, illustrates that a modest increase in average score can be practically important.

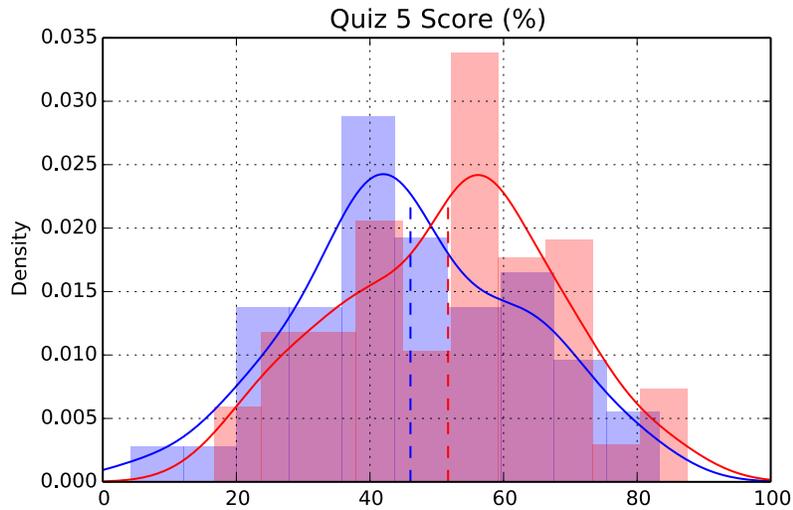

**Figure 2.** Distribution of scores for the control and treatment groups on quiz 5 in the winter quarter. The dashed vertical lines designate the means. (The difference in means on this quiz was 5.9.) Similar plots for all of the quizzes and final exam may be found in the supplement.

To interpret this effect size in context, we examined the "effect sizes" of well-known correlates of achievement, such as gender and race, on quiz scores (Table 1). Note that, unlike peer assessment, these are not randomized interventions, so these numbers should not be interpreted as causal effects; they are included only for comparison. We see that the effect of peer assessment represents about 40% of the gender achievement gap and about 20% of the racial achievement gap, which are persistent challenges in college science classes.[19,20] Furthermore, the effect of peer assessment is about 25% of the effect of having previously taken a statistics class.

| Factor | Effect Size |
| --- | --- |
| Peer assessment (short term) | .11 (.04) |
| Peer assessment (long term) | .12 (.04) |
| Gender achievement gap (1=male, 0=female) | .32 (.12) |
| Racial achievement gap (1=underrepresented minority) | −.60 (.14) |
| Statistics background (1=passed AP stats) | .54 (.11) |
| Math background | .68 |
| (1=course beyond calculus) | (.10) |
| Class year (1=upperclassman) | .25 (.12) |

**Table 1.** The effect sizes of various correlates of achievement, as compared with the effect size of peer assessment. (Standard errors are shown in parentheses.)

**Discussion:**

This study has established that peer assessment produces concrete gains in student achievement, above and beyond the effect of receiving feedback. Thus, peer assessment is unusual among educational interventions in that the traditional cost-benefit tradeoff seems not to apply: it helps students learn *and* saves instructors time. This suggests that peer assessment is a valuable educational tool that belongs in all classes, not just MOOCs and large classes where it is the only option.

In addition, we have shown the role that online resources can play in education research, echoing a recent development in the MOOC literature.[21] These resources make individual-level randomizations, the core of high quality RCTs, feasible. While we have focused on peer assessment specifically, a similar study design could be used to investigate other questions as well. The truly revolutionary impact of web-based educational tools may be the ability to apply the scientific method efficiently inside the classroom, enabling unprecedented insight into the learning process.

**Acknowledgements:**

We are grateful to Sergio Bacallado and Renjie You for helpful discussions, to Cyrus DiCiccio, Zhou Fan, Scott Freeman, Max Grazier-G'sell, Trey Wright and Carl Wieman for their careful reading of our paper and helpful suggestions, and to Daphne Koller for encouragement. This project was supported by a seed grant from the Vice Provost for Online Learning, Stanford University.

# Supplementary Information for "Peer assessment enhances student learning"

Dennis L. Sun    Naftali Harris    Guenther Walther    Michael Baiocchi

October 14, 2014

## 1 Materials and Subjects

### 1.1 Course Structure

Stats 60 (Introduction to Statistical Methods) is an introductory, pre-calculus statistics course at Stanford University. It is offered every academic term, with four lectures and one recitation section per week. The class fulfills the math general education requirement for undergraduates, the statistics requirement for pre-medical students, and the statistics requirement for psychology majors. It is one of the largest courses at Stanford, taken by a diverse group of students.

The course follows the textbook by Freedman, Pisani, and Purves (*1*). The syllabus of the ten-week course is shown in Table 1. We divided the curriculum into one introductory unit (Unit 1) and four main units (Units 2-5). The first unit covers material that is central to the remaining units (e.g., the normal distribution), but the other units are mostly self-contained, meaning that material covered in one unit is roughly independent of the material in another. This structure allowed us to run a crossover study, described in Section 2.1.

Students were required to submit weekly homeworks through an online homework system, OHMS, designed for this course (*2*). Homeworks consisted of some questions that were automatically scored by the machine (e.g., multiple choice and questions with numeric answers), as well as free-response questions. The peer assessment treatment was applied only to the free-response questions, and students were allowed to participate in peer assessment only if they had answered the corresponding question. Table 2 shows the timing of the homework submissions and grading periods each week. Both peer and instructor graders adhered to the same grading period, and feedback was released to students at the conclusion of the grading period.

In the two terms that we ran the study, each unit was concluded by a unit quiz, which was administered on the Wednesday following the conclusion of the unit. This ensured that students had received feedback on all homework before taking the unit quiz. In addition, a final exam was administered in Week 11. All assessments were graded by members of the course staff who were blinded to the treatment assignments. Each question was graded by a single person to ensure consistency.

| Unit | Week | Material | Assessment |
|---|---|---|---|
| 1 | 1 | histograms, summary statistics, normal curve | |
| 2 | 2 | experimental design, correlation | Unit 1 Quiz |
|   | 3 | regression | |
| 3 | 4 | probability | Unit 2 Quiz |
|   | 5 | expected value, standard error | |
| 4 | 6 | sample surveys | Unit 3 Quiz |
|   | 7 | confidence intervals | |
| 5 | 8 | one-sample hypothesis tests | Unit 4 Quiz |
|   | 9 | two-sample tests and chi-square tests | |
| — | 10 | [review] | Unit 5 Quiz |

Table 1: Course syllabus, showing the topics covered in each unit, along with

| Su | M | Tu | W | Th | F | Sa |
|---|---|---|---|---|---|---|
| | | |——homework——||–peer | | | |
| assess.—|| | |——homework——||–peer | | | |
| assess.—| | | quiz | | | |

Table 2: Schedule for each two-week unit. Each week was divided into homework and peer assessment periods. Note that the material for the unit was covered in the two weeks, but the peer assessment and quiz spill over into the following week, after the next unit has begun.

The five unit quizzes together comprised 40% of the overall course grade, while the homework and peer assessment accounted for 10% each. The final exam represented the remaining 40%. No scores were dropped in the computation of the course grade, incentivizing students to do well on every component.

## 1.2 Students

In the autumn quarter, 150 students were enrolled in Stats 60 for credit and responded to the study consent form. (The consent form was included as part of the regular, weekly online homework and was required. The students who did not respond either joined the class late or eventually withdrew from the class.) Of these 150 students, 2 students opted out of the study. In the winter quarter, 240 students took the course for credit and responded to the study consent form. Of these 240 students, 1 student opted out of the study.

|                              |            | Autumn       | Winter       |
|------------------------------|------------|--------------|--------------|
| Underrepresented minority    |            | 39 (26.3%)   | 69 (28.9%)   |
| Gender (male)                |            | 53 (35.8%)   | 95 (39.7%)   |
| Previous statistics course   |            | 30 (20.3%)   | 43 (18.0%)   |
| Advanced math course         |            | 54 (36.5%)   | 85 (35.6%)   |
| Class year                   | Freshman   | 26 (17.6%)   | 41 (17.2%)   |
|                              | Sophomore  | 40 (27.0%)   | 50 (20.9%)   |
|                              | Junior     | 37 (25.0%)   | 67 (28.0%)   |
|                              | Senior     | 41 (27.7%)   | 76 (31.8%)   |
|                              | Graduate   | 4 (2.7%)     | 5 (2.1%)     |
| Total                        |            | 148          | 239          |

Table 3: Covariates for the 387 students. "Advanced math course" is defined as a course in multivariable calculus or linear algebra. Although the course requires no calculus, more than 90% of students in both quarters had taken single-variable calculus.

Covariate data for the 387 (148 + 239) subjects is shown in Table 3. The demographics across the two quarters are fairly similar.

## 1.3 Homeworks and assessments

Each week's homework consisted of six multiple choice and numeric answer questions that were automatically scored by computer, as well as three free-response questions that were graded by a human (either a peer grader or an instructor). Each unit quiz consisted of six questions, all free-response, which tested conceptual understanding or problem solving. The final exam was longer and comprehensive, but the questions were otherwise similar to the quizzes. The same homeworks were used in both quarters, although different assessments were used.

Assessment questions were similar in format to, and tested the same concepts as, homework questions. However, care was taken to ensure that no question simply repeated a homework question to discourage students from memorizing questions instead of internalizing the material. (Although repeating a homework question verbatim on a quiz might inflate the estimated effect for one unit, this strategy would backfire because of the crossover design, since all students would exploit this in subsequent weeks, nullifying any treatment effect.)

Figure 1 shows a question that appeared on the homework and a corresponding quiz question that tested the same concept. Memorizing the answer to the homework problem would not help with the unfamiliar quiz question, but internalizing the concept in the homework problem.

**Homework (Free-Response) Question**

> A box contains one red marble and nine green ones. Five draws are made at random with replacement. The chance that exactly two draws will be red is given by the binomial formula:
>
> $$\frac{5!}{2!3!} \times \left(\frac{1}{10}\right)^2 \left(\frac{9}{10}\right)^3.$$
>
> Is the addition rule used in deriving this formula? Answer yes or no, and explain carefully.

**Corresponding Quiz Question**

> A standard deck of 52 cards has 13 clubs. You bet your friend Ben that there will be exactly 3 clubs among the first 6 cards drawn. Suppose that the cards are flipped over one at a time, without replacement. What's the chance you win? (Hint: First, try calculating the chance of a particular ordering of clubs and no clubs.)

Figure 1: An example question from the homework and a question from the corresponding unit quiz. Although one question is conceptual and the other is computational, they test the same idea: every sequence has the same probability, with or without replacement, so to calculate the chance of 2 total reds out of 5 draws (or 3 total clubs out of 6 draws), one computes the probability of a particular sequence and then multiplies by the number of combinations.

## 2 Methods

### 2.1 Experimental design

The goal of the study was to understand the effect that participating in peer assessment has on achievement. The introductory unit (Unit 1) was excluded from the study, so that all students had a chance to become familiar with peer review before the start of the study. As a side benefit, this allowed us to use the scores on the Unit 1 quiz as a baseline measure of student ability. We conducted a crossover study on the four remaining units: each student was assigned to treatment during exactly two of the units and to control during the other two. This within-subject design not only enhanced the power of the study to detect small effect sizes, but also served a logistical purpose, ensuring that each student had the same total workload for the course.

A crossover design ensures covariate balance, since each subject serves as his or her own control. However, since the measurement instruments (e.g., quizzes) may differ from unit to unit, randomization is also necessary to ensure that the treatment and control groups are balanced within each unit in order to account for these differences. For example, if treatment assignments are correlated with the difficulty of the quizzes, then difficulty becomes a confounding variable. However, complete randomization only guarantees balance *on average*, so we used matched pairs randomization in order to provide a stronger safeguard against potential imbalance.

|       | Unit |   |   |   |
|-------|------|---|---|---|
| Group | 2    | 3 | 4 | 5 |
| 0     | T    | C | T | C |
| 1     | C    | T | C | T |
| 2     | T    | C | C | T |
| 3     | C    | T | T | C |

Table 4: Treatment assignments for the four treatment groups. T indicates treatment, and C indicates control.

|              | Autumn              | Winter              |
|--------------|---------------------|---------------------|
| Unit 1 Quiz  | $F(3, 120) = .17$   | $F(3, 190) = .24$   |
|              | $p = .92$           | $p = .87$           |
| Class Year   | $\chi^2(9) = 6.4$   | $\chi^2(9) = 7.7$   |
|              | $p = .70$           | $p = .56$           |
| AP Statistics| $\chi^2(3) = .96$   | $\chi^2(3) = 1.7$   |
|              | $p = .81$           | $p = .64$           |

Table 5: Tests of covariate balance between the four treatment arms. The p-values indicate that the four treatment arms are not significantly different.

In the matched pairs randomization, students were first blocked into groups based on gender, race, and previous statistics background. Then, within each block, each student was paired with the "most similar" student using optimal non-bipartite matching on covariates such as Unit 1 quiz score, class year, and math background (*3*). Then, within each pair, a coin was flipped to assign the members to complementary treatment groups. Each pair was either assigned to groups 0/1 or groups 2/3 (see Table 4 for the definitions of the groups). This design eliminates the possibility of drastic covariate imbalance, since each student in the treatment group is balanced by a similar student in the control group in all four units of the study. See Table 5 for an assessment of the balance for some of the baseline covariates.

## 2.2 Effect size estimation

In educational research, the measurement instruments (i.e., exams) vary as to difficulty and disciminatory power. These two factors are precisely the ones captured by item-response theory (4). The essence of the IRT model is that student $i$'s score on exam $j$, conditional on his or her ability $\theta_i$, is

$$\mathbb{E}(Y_{ij}|\theta_i) = \sigma_j \theta_i + \mu_j.$$

Note that this differs from standard ANOVA or random effects models only in the introduction of a exam-dependent variance, $\sigma_j^2$. By recognizing that exams not only vary in difficulty (i.e., mean) but also in discriminatory power (i.e. variance), we obtain a more nuanced model of exam scores.

Although $\theta_i$ represents a student's latent ability, there may be variations in one's performance on a given day; this is captured by a noise term $\epsilon_{ij}$. Therefore, a more explicit representation of the exam scores $Y_{ij}$ is:

$$Y_{ij} = \sigma_j(\theta_i + \epsilon_{ij}) + \mu_j \tag{1}$$

where for identifiability, we assume $\mathbb{E}(\theta_i) = \mathbb{E}(\epsilon_{ij}) = 0$ and $\text{Var}(\theta_i) + \text{Var}(\epsilon_{ij}) = 1$. We do not make any distributional assumptions about $\theta_i$ and $\epsilon_{ij}$. Under these assumptions, we have:

$$\mathbb{E}(Y_{ij}) = \mu_j \qquad\qquad \text{Var}(Y_{ij}) = \sigma_j^2 \tag{2}$$

Now suppose that we introduce an intervention. Because exam scores have no inherent meaning, the raw effect size (i.e., the difference in average scores between the treatment and control groups) is not meaningful. For example, if every question on an exam were worth twice as much, then the raw effect size would double. Standard practice is to report a *standardized* effect size, i.e., express the effect size in terms of standard deviations (5). This allows researchers to aggregate effect sizes across studies.

The implicit assumption that underlies this practice is that the raw effect is a constant multiple of the exam standard deviation, i.e., $\tau \sigma_j$. By standardizing this raw effect by the standard deviation, one obtains an estimable quantity $\tau$ that is constant across exams.

To summarize this discussion, if student $i$ is randomly assigned to treatment $W_{ij} \in \{0,1\}$ on exam $j$, then his or her score $Y_{ij}$ is modeled by:

$$Y_{ij} = \sigma_j(\theta_i + \tau W_{ij} + \epsilon_{ij}) + \mu_j. \tag{3}$$

We propose the following procedure for estimating the effect size $\tau$:

- Standardize the scores on each exam by the observed mean $\hat{\mu}_j$ and standard deviation $\hat{\sigma}_j$ of the scores of students in the control group:

$$\hat{Z}_{ij} := \frac{Y_{ij} - \hat{\mu}_j}{\hat{\sigma}_j}$$

- For student $i$, compute the difference $D_i$ between the average ($z$-)score when he or she was assigned to treatment and the average ($z$-)score when assigned to control.

- The average of the difference, $\bar{D}$, estimates the effect size.

To understand why this is an estimate of the effect size, we note that $\hat{\mu}_j$ and $\hat{\sigma}_j$ are consistent estimators for $\mu_j$ and $\sigma_j$. Therefore, we can consider $Z_{ij} = \frac{Y_{ij} - \mu_j}{\sigma_j}$ instead of $\hat{Z}_{ij}$ without any loss of generality. We thus obtain:

$$Z_{ij} = \theta_i + \tau W_{ij} + \epsilon_{ij}.$$

Now the difference $D_i$ for individual $i$ is:

$$D_i = \frac{1}{m/2} \sum_{j=1}^{m} (2W_{ij} - 1) Z_{ij}$$

$$= \frac{1}{m/2} \sum_{j=1}^{m} (2W_{ij} - 1)(\theta_i + \tau W_{ij} + \epsilon_{ij})$$

Now we apply the identities $\sum_{j=1}^{m} W_{ij} = m/2$ (since the design is balanced) and $W_{ij}(2W_{ij} - 1) = W_{ij}$ to obtain:

$$= \tau + \frac{1}{m/2} \sum_{j=1}^{m} (2W_{ij} - 1) \epsilon_{ij}.$$

Next, we establish that the $D_i$ are independent and identically distributed. First, conditional on the treatment assignments $W := (W_{ij}), i = 1, ..., n, j = 1, ..., m$, the second term is equal in distribution to any fixed permutation of the assignments:

$$(D_1, ..., D_n) \mid W \stackrel{d}{=} \left( \tau + \frac{1}{m/2} \left[ \sum_{j=1}^{m/2} \epsilon_{1j} - \sum_{j=m/2+1}^{m} \epsilon_{1j} \right], ..., \tau + \frac{1}{m/2} \left[ \sum_{j=1}^{m/2} \epsilon_{nj} - \sum_{j=m/2+1}^{m} \epsilon_{nj} \right] \right).$$

Examining the right-hand side, we see that the $D_i$ given $W$ are i.i.d. and, moreover, the conditional distribution does not depend on $W$, so the $D_i$ are also i.i.d. unconditionally.

Therefore, we can apply the Central Limit Theorem to obtain

$$\sqrt{n}(\bar{D} - \tau) \Rightarrow N(0, \cdot). \tag{4}$$

This establishes both the consistency of the estimator $\bar{D}$ for estimating $\tau$, as well its asymptotic normality.

## 2.3 Significance Testing

There are two approaches to significance testing. One is to use the asymptotic normality result (4), and conduct a $z$ or a $t$ test of the hypothesis $\tau = 0$. Although we have assumed that the treatment effect is constant for all individuals, i.e., $\tau_i = \tau$, the above procedure in fact controls Type I error under the less stringent hypothesis $\mathbb{E}(\tau_i) = 0$.

An alternative approach, which does not depend on the validity on the model described in Section 2.2, is a permutation test. Although permutation tests are nonparametric, they in general test the sharp null hypothesis of zero treatment effect for all individuals, i.e., $\tau_i = 0$.

We found that the normal approximation to be so accurate on our data (see Figure 4) that there is essentially no difference between the two approaches. The standard errors and $p$-values, as calculated using the asymptotic approximation and from the permutation distribution, are the same.

## 2.4 Combining data sources

Our study ran for two academic terms. One appeal of the above approach is that the differences $D_i$ already account for differences between students and exams in the two quarters. Therefore, we can simply combine the differences $D_i$ from the two terms into one dataset. This produces a tremendously powerful procedure, as evidenced by the small $p$-values despite the moderate effect size.

# 3 Results

## 3.1 Key findings

The effect sizes of various determinants of outcome are summarized in Table 6. The first column shows the aggregate effect (which is a reproduction of Table 1 from the main paper). The second and third columns show the estimated effect in each quarter. The short term effect of peer assessment was calculated from the unit quizzes, using the permutation method described in Section 2.2. The long term effect was calculated in the same way from the final exams.

The effect sizes for the other determinants of achievement were calculated by taking the standardized difference in average Unit 1 quiz score between the relevant groups. Because Unit 1 was not included in the study, the Unit 1 quiz scores are not contaminated by the intervention. However, one should be careful about ascribing a causal interpretation to these effects. For example, the effect of a previous statistics course (AP Stats) was found to be .54, but this is potentially confounded with other factors, since we simply observed who had taken AP Statistics and who had not. In fact, it may even be confounded with the other determinants that we are examining, such as gender, race and math background.

| Factor | Effect (Overall) | Effect (Autumn) | Effect (Winter) |
|---|---|---|---|
| Peer assessment (short term) | .11 (.04) | .08 (.06) | .14 (.05) |
| Peer assessment (longer term) | .12 (.04) | .12 (.06) | .12 (.05) |
| Gender achievement gap (1 = Male) | .32 (.12) | .16 (.19) | .41 (.15) |
| Racial achievement gap (1 = URM) | −.60 (.14) | −.65 (.23) | −.57 (.17) |
| Statistics background (1 = Passed AP Stats) | .54 (.11) | .54 (.16) | .55 (.16) |
| Math background (1 = Course beyond calculus) | .68 (.10) | .67 (.17) | .69 (.13) |
| Class Year (1 = Upperclassmen) | .25 (.12) | .12 (.18) | .33 (.15) |

Table 6: Summary of effect sizes, broken down by quarter

Therefore, the .54 should not be interpreted as "the effect of taking AP Statistics" but rather just a benchmark against which the effect size of peer assessment can be compared.

We also show the distribution of quiz scores for each quiz in Figures 2 and 3. These figures demonstrate the large amount of variability present in student scores. They also reveal that the effect of the treatment is quite subtle. However, the crossover design allows us to detect such small effects with high precision. One way to tell that the treatment effect is positive from these figures by noting that the treatment group is never worse, but it is sometimes very clearly better, as in the case of Quiz 2 in autumn quarter or Quiz 5 in winter quarter.

Finally, we examine the permutation distribution of the observed effect, under the null hypothesis of no effect. In Figure 4, we see that the observed effect is far in the tails of the permutation distribution and highly statistically significant.

## 3.2 Issues of Compliance

As with any study, not all students complied with their treatment assignment. Each student was assigned to grade 3 questions in each of the 4 weeks they were assigned to the treatment, for a total of 12 questions during the quarter. Figure 5 shows the distribution of questions completed. Although the vast majority of the class completed all of the peer assessment assignments, some students completed fewer, with a handful completing no peer assessments at all. (The bumps at 9 and 6 indicates that students tended to miss entire weeks of peer assessment when they missed questions at all.)

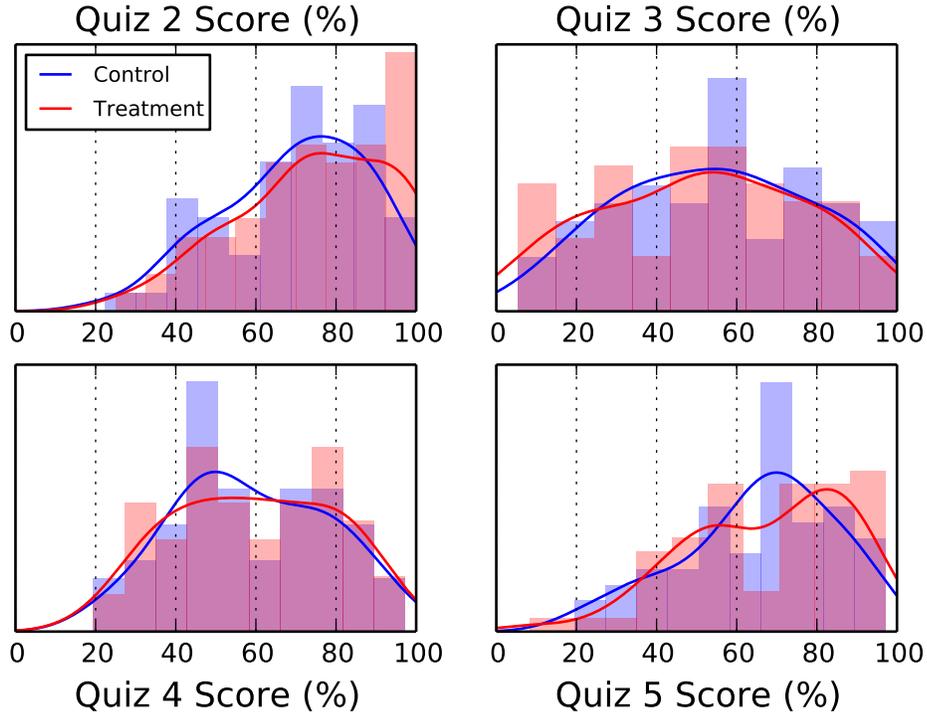

Figure 2: Distribution of quiz scores in autumn quarter, along with an estimate of the density. Red is the treatment group, while blue is the control group.

The first problem is how to handle students who did not comply with the treatment. In an intent-to-treat analysis, one only considers the treatment that was assigned. However, we argue that in educational interventions, the estimand of interest is typically the effect on students who would comply with a treatment if required. The effect on students who would not receive the treatment either way can be assumed to be close to zero. Because non-compliers are systematically excluded from both the treatment and control groups, we are able to obtain an unbiased estimate of the effect size on the subpopulation. This type of analysis is typical in the education literature. For example, Miyake et al. examined the effect of an writing exercise on student outcomes; they focused only on the 399 students (out of 602) who attended class and participated in the exercise (*6*).

The second problem is how to define compliance. In the above analysis, we defined compliance as completing at least 10 of the 12 peer assessment questions assigned. This definition resulted in about 83% compliance, leaving us with a final sample of 322 students. Both the short-term and long-term effect sizes of the treatment on the non-compliers were about zero ($t = .44$, $p = .66$), as expected.

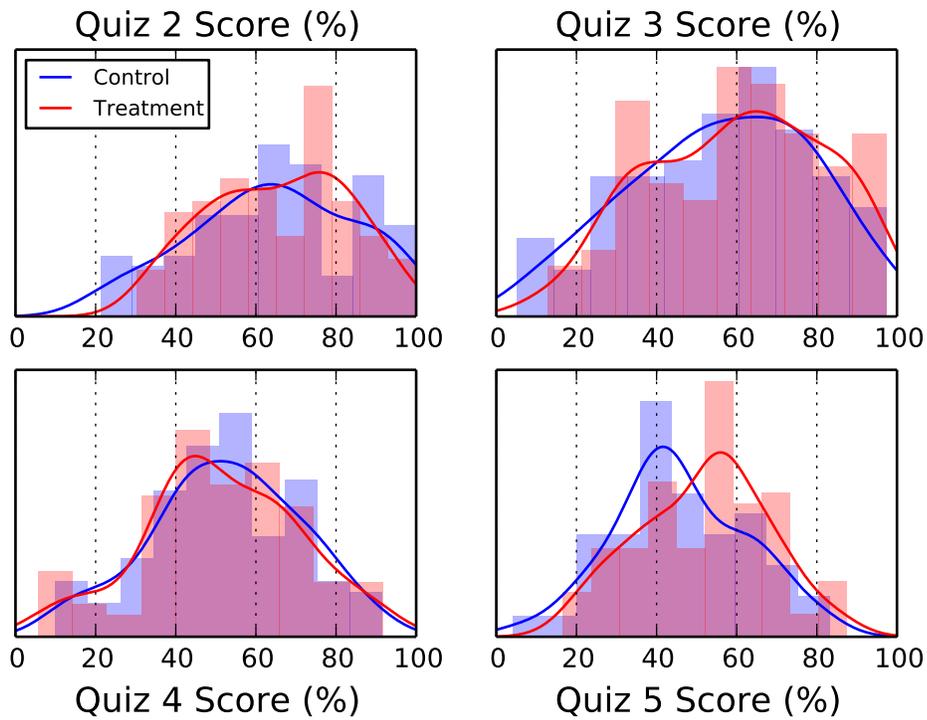

Figure 3: Distribution of quiz scores in winter quarter, along with an estimate of the density. Red is the treatment group, while blue is the control group.

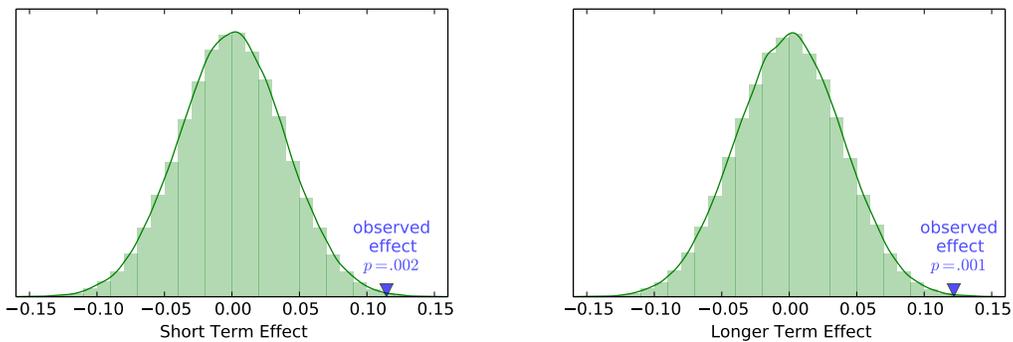

Figure 4: The permutation distribution of the test statistic $\bar{D}$ on the quizzes and on the final exam, under the null hypothesis of no effect. The observed statistic is quite extreme.

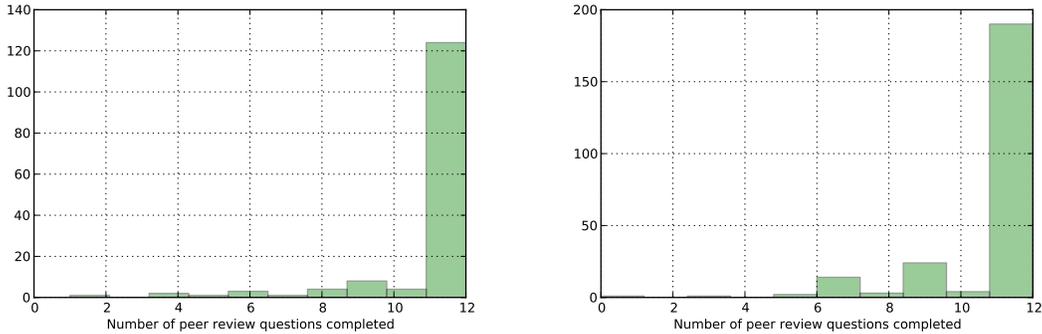

Figure 5: Histograms showing the distribution of number of peer assessment questions completed in autumn quarter (left) and winter quarter (right).

However, one could consider other definitions as well. We now present an analysis of the sensitivity of our results to the compliance definition. If we had instead defined compliance as completing all 12 questions, the compliance would drop to about 75%; the estimated short-term effect would be slightly smaller at $d = .10$, but the longer-term effect would be larger at $d = .13$. On the other hand, if we had instead defined compliance as completing at least 8 questions, the compliance rate would be 93%, while the short-term and longer-term effects would both be $d = .10$. Therefore, we see that our findings are fairly robust to how compliance is defined.

### 3.3 Testing for Quarter and Carryover Effects

Our analysis rests on two primary assumptions. First, by aggregating the data from the two quarters into one, we are assuming that there is no interaction between treatment and quarter. We already saw in Table 3 that there is essentially no difference in the student demographics between the two quarters. However, there may still be an instructor effect, since different instructors taught the two quarters. However, because there was no difference in the average effect size between quarters ($t = .72, p = .47$), we conclude that there is no quarter effect whatsoever.

Second, by using the unit quiz scores as a measure of learning in that unit, we are assuming that there are no carryover effects, i.e., the effect of a treatment in Unit 2 does not "carry over" into Unit 3. Although this assumption is often difficult to test, we use the following heuristic: if there were carryover effects from one unit to another, then we would anticipate different benefits accruing to different treatment groups (e.g., students assigned to TCTC might show more benefit than CTTC). However, we found that there was no difference between the four treatment groups ($F(1, 294) = 0.61$, $p = .43$), which is consistent with the assumption of no carryover effect.

## 3.4 Heterogenous Treatment Effects

Students may respond differentially to pedagogical techniques. To investigate whether socio-economic factors were associated with differential treatment effects we ran a multivariable regression model with gender, self-identified race, unit 1 quiz scores, prior statistics coursework, and class year. The outcome variable was the estimated individual-level treatment effect, as calculated using the methods in Section 2.2. This represents a "blunt" method for detecting heterogeneity in the treatment effect.

We found no connection between the covariates and the estimated individual-level treatment effect ($F(14, 277) = 0.67$, $p = 0.81$). Additionally, we fit marginal models that focused on gender only ($F(1, 294) = .01$, $p = .94$) and race only ($F(1, 294) = .20$, $p = .66$), confirming that the effect seems to be relatively homogenous across these socioeconomic factors.

It is important to note that the discussion in this subsection is about heterogeneity of the treatment effect. That is, we are looking to see if there are subgroups which benefit more or less from the treatment than other subgroups. This is not to be confused with relative performance on the quizzes across subgroups. As noted in Section 3.1, there were substantial performance differences between racial and gender subgroups.

Using scores on quiz 1 as a proxy for student's aptitude, we modeled the individual treatment effect to explore the possibility that the treatment effect may differentially impact students. One hypothesis was that that the most and least skilled students would not benefit much from peer assessment, and that it would produce a benefit only for those in the middle. We ran a linear model of estimated individual-level treatment effect on quiz 1 which produced an insignificant connection ($F(1, 292) = 1.6$, $p = .20$). To assess the possibility of a non-linear connection we fit a loess curve to the data. Both models are shown in Figure 6.

## 3.5 Spillover effects

A common assumption in most analyses is the absence of spillover effects. For example, in this study, if someone on the treatment wing of the study gained a better understanding of p-values from assessinging her peers' answers and then transferred this insight to a friend on the control wing, this would be an example of spillover which could potentially bias our estimates of the treatment effect. This is a common challenge in studies of educational outcomes and is formally referred to as a violation of the stable unit treatment value assumption (SUTVA).

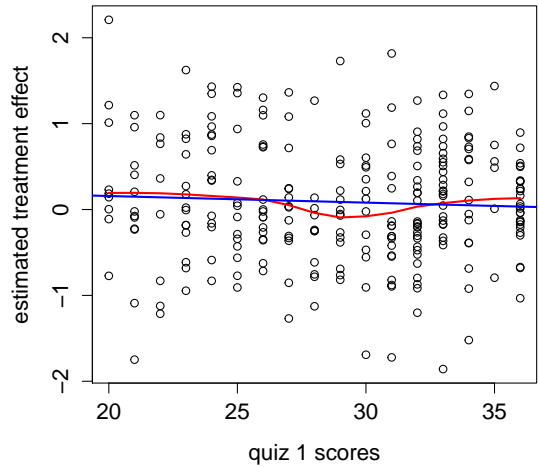

Figure 6: Scatterplot of individual-level treatment effect (y-axis) and quiz 1 scores (x-axis). Quiz 1 is a proxy for student aptitude for the course. The blue line is the least squares line, while the red line shows a nonparametric model fit to the same data.

Because we did not restrict students from interacting and helping one another, there may be spillover effects. However, spillover effects would bias the effect towards zero, since the transfer of information would tend to make students more similar. This means that our findings may actually be conservative—that the true effect of peer assessment is even larger.

### 3.6 Student attitudes and behaviors

Peer assessment demands additional time from students, so it must be limited in scope in order to be feasible. Using server logs, we were able to obtain an estimate of how long each student spent on homework. Although it is difficult to track how long each student spent on peer assessment exactly, we obtained a rough estimate by examining the difference in timestamps between successive submissions. We filtered out any differences that were longer than one hour (suggesting that the student had left his or her computer and returned to it later). It is possible to back out how long students spent on peer assessment from these differences.

Figure 7 shows a scatterplot of the amount of time each student reported spending and the time they actually spent, as estimated using the above procedure. We see that students tended to overestimate the amount of time they spent; students reported spending around 35 minutes, whereas the actual number was about 27 minutes. Compared to the 10 hours students spend on the course per week, the additional 20-30 minutes that is required for peer assessment is negligible.

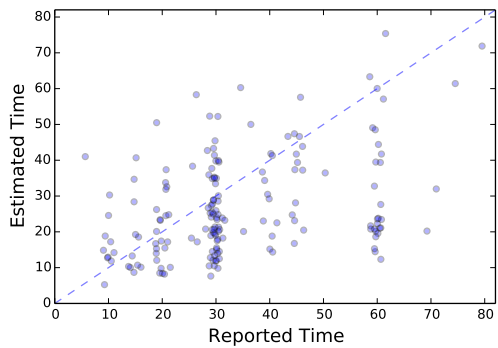

Figure 7: Scatterplot of the amount of time each student reported spending on peer assessment per week, versus the time they actually spent (as estimated from the server logs). The points have been jittered to distinguish them.

We also surveyed students on their perception of the benefit of peer assessment on a scale from 1 to 5, with 1 indicating "not helpful at all" and 5 indicating "extremely helpful". Although the median student reported finding peer assessment only "somewhat helpful," there was virtually zero correlation ($r = .01$, $p = .94$) between a student's perception of the benefit and the estimated benefit. This affirms our concern that surveys may not be the best measure of student learning. The full data is shown in Figure 8. The benefit is very clearly positive overall, but there does not appear to be any relationship between the perceived and estimated benefits.

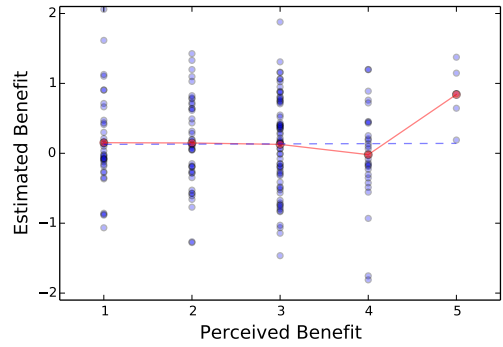

Figure 8: Scatterplot of the estimated benefit of peer assessment for each student, versus the perceived benefit (on a scale from 1-5). The red curve traces the means in each category, while the blue dotted line denotes the linear regression line.

6. Miyake A, et al. (2010) Reducing the gender achievement gap in college science: A classroom study of values affirmation. *Science* 330:1234–1237.